\documentclass[acmtog,authorversion]{acmart}

\usepackage{booktabs} 

\acmPrice{15.00}

\setcopyright{acmlicensed}
\acmJournal{TOG}
\acmYear{2018}
\acmVolume{37}
\acmNumber{6}
\acmArticle{270}
\acmMonth{11}
\acmDOI{10.1145/3272127.3275104}

\setcitestyle{square}

\usepackage{color}
\newcommand{\red}[1]{#1}
\newcommand{\green}[1]{#1}
\newcommand{\rev}[1]{#1}
\newcommand{\revb}[1]{#1}
\newcommand{\revc}[1]{#1}
\newcommand{\revd}[1]{#1}

\begin{document}

\title[Relighting Humans]{Relighting Humans: Occlusion-Aware Inverse Rendering for Full-Body Human Images}


\author{Yoshihiro Kanamori}
\affiliation{%
  \institution{University of Tsukuba}}
\email{kanamori@cs.tsukuba.ac.jp}

\author{Yuki Endo}
\affiliation{%
  \institution{University of Tsukuba \& Toyohashi University of Technology}}
\email{endo@val.cs.tut.ac.jp}
%
%

\renewcommand{\shortauthors}{Kanamori and Endo}

\begin{abstract}
Relighting of human images has various applications in image synthesis. For relighting, we must infer albedo, shape, and illumination from a human portrait. Previous techniques rely on human faces for this inference, \revb{based on} spherical harmonics (SH) lighting. However, because they often ignore light occlusion, inferred shapes are biased and relit images are unnaturally bright particularly at hollowed regions such as armpits, crotches, or garment wrinkles. This paper introduces the first attempt to infer light occlusion in the SH formulation \revb{directly}. Based on supervised learning using convolutional neural networks (CNNs), we infer not only an albedo map, illumination but also a {\em light transport map} that encodes occlusion as nine SH coefficients per pixel. The main difficulty in this inference is the lack of training datasets compared to unlimited variations of human portraits. Surprisingly, geometric information including occlusion can be inferred plausibly even with a small dataset of synthesized human figures, by carefully preparing the dataset so that the CNNs can exploit the data coherency. Our method accomplishes more realistic relighting than the occlusion-ignored formulation. 
\end{abstract}

\ccsdesc[500]{Computing methodologies~Rendering}
\ccsdesc[500]{Computing methodologies~Image-based rendering}

\keywords{inverse rendering, light transport, convolutional neural network}

\begin{teaserfigure}
 \centering
 \includegraphics[width=\linewidth]{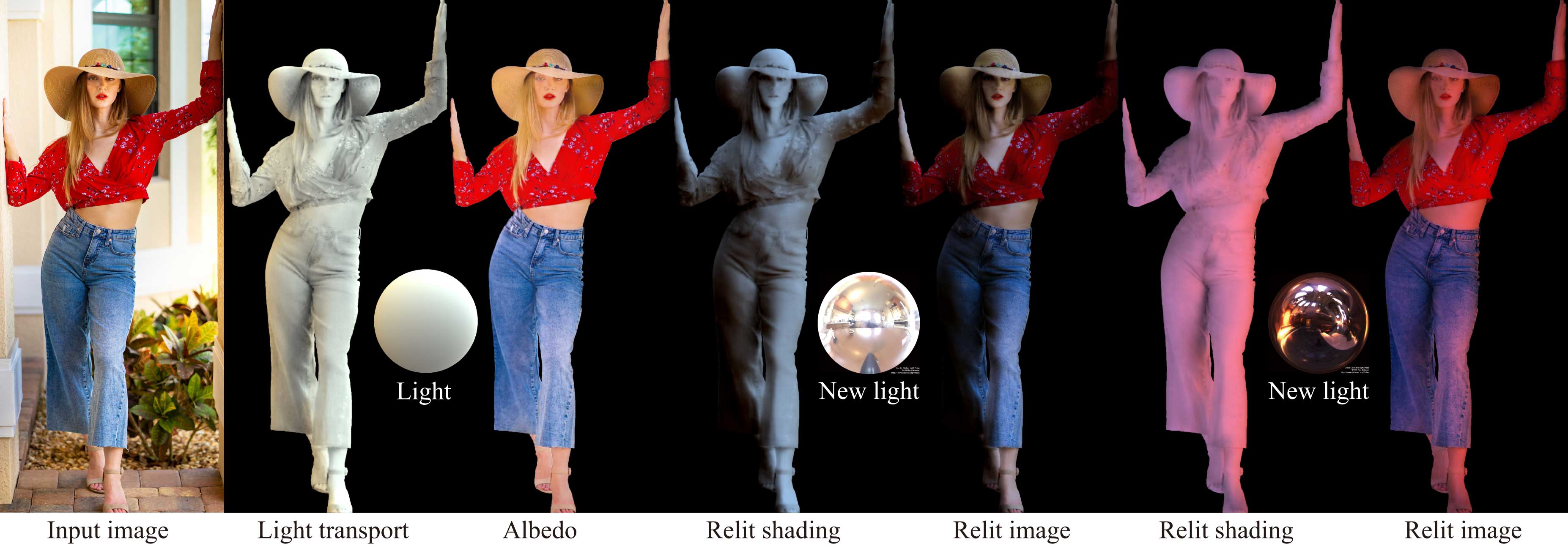}
 \caption{Given a single human image and its mask, our method infers \revb{a {\em light transport map}} (i.e., coefficient vectors of \revb{second}-order spherical harmonics that encode light occlusion), \revb{light, and albedo map} using convolutional neural networks, and allows \revb{fast} relighting of the human figure under different illuminations with self-shadows. Input image courtesy of Jose Martinez.}
  \label{fig:Teaser}
\end{teaserfigure}

\maketitle

\section{Introduction}
\label{sec:Introduction}

Relighting of human images has various applications in image \revb{synthesis} such as stylized shading of portraits~\cite{DBLP:journals/tog/ChaiLSCHZ15,DBLP:journals/tog/ShuHSSPS17} or cut \& paste of human image clips~\cite{DBLP:journals/tog/XueADR12}. 
For \rev{physically}-based relighting of a human portrait, we must infer reflectance, \revb{shape,} and illumination from the single image.
\rev{Previous} techniques obtain the cues of albedo and shape from human faces via fitting of morphable 3D face models~\cite{DBLP:conf/siggraph/BlanzV99} or inference based on convolutional neural networks (CNNs)~\cite{DBLP:conf/CVPR2018/abs-1712-01261}, and infer illumination on the basis of {\em spherical harmonics} (SH) lighting~\cite{DBLP:conf/siggraph/RamamoorthiH01a,1177153}.

The SH-based \red{lighting} yields an elegant analytical formulation of \red{shading from surface normals and illumination if} \revb{we ignore the} light occlusion; \revb{we can calculate} per-pixel SH bases from \red{normals}, and then illumination is obtained in the form of SH coefficients using least squares~\cite{DBLP:journals/pami/Kemelmacher-ShlizermanB11}.
However, as is well known in the realtime-rendering literature, rendered images without light occlusion lacks photorealism because hollowed regions become unnaturally bright, \revd{although they ought to be occluded, compared to other regions.} 
An approximate solution is to darken the hollowed regions by multiplying scalar values depending on occlusion, i.e., {\em ambient occlusion}~\cite{DBLP:conf/rt/ZhukovIK98}. 
\rev{A} more elegant solution is to encode light occlusion and cosine decay as SH coefficients, which we refer to as a {\em light transport vector}, and formulate lighting calculation as a dot product of the light transport vector and SH coefficients of illumination~\cite{DBLP:journals/tog/SloanKS02}.
Unfortunately, calculating occlusion requires the geometry to be inferred, and is quite computationally expensive due to visibility sampling at each surface point.

In this paper, we introduce the first attempt to infer not only diffuse albedo but also a light transport vector \red{for} each pixel from a masked full-body human image, \revb{which} is accomplished by supervised learning using CNNs, with a ground-truth training dataset synthesized from scanned 3D human figures.
The main problem \revd{in making} this inference possible is the lack of \revb{a} training dataset, considering the unlimited variations of human portraits \revb{regarding} poses, genders, builds, and garments. \revb{To} the best of our knowledge, there is only a single publicly-available dataset of scanned 3D human figures~\cite{DBLP:conf/cvpr/ZhangPBP17}, but it lacks variations (i.e., only \red{five} individuals with one or two \revd{outfits each}). 
We additionally purchased commercial datasets of clothed 3D human figures, amounting to only a few hundreds of models.
Surprisingly, \red{by carefully selecting standing figures and aligning them in the training images}, CNNs can learn plausible light transport \red{vectors}, which can capture occlusions at armpits, crotches or garment wrinkles, even from such a small dataset.
This \revb{result} implies that CNNs can learn geometric information including occlusion from the silhouettes of human figures, i.e., \red{binary masks} to some extent, which is a similar conclusion drawn from \red{the recent} work inferring normal maps only from silhouette lines~\cite{Lun:2017:SketchModeling}.

Thanks to the inferred light transport maps, we can relight human portraits quite efficiently just by calculating dot products of light transport vectors and SH coefficients of light, followed by channel-wise multiplication of inferred albedo maps.
The inference of albedo and light transport maps is fast (0.43 sec. for each $1024\times1024$ image), and our inferred albedo and light transport \red{vectors} have sufficient quality for plausible relighting of human images, as shown in Figure~\ref{fig:Teaser}.

\section{Related Work}
\label{sec:RelatedWork}

For single-image physically-based relighting, we must solve {\em inverse rendering}, i.e., estimation of shape, reflectance, and illumination from a single image, \revb{which} is a highly ill-posed problem.
Classical methods relax it by assuming that some of the three components are known, or use prior knowledge of the target in order to estimate the remaining components. 
Recent methods adopt data-driven approaches that exploit statistics of the three components in the target domain.

\paragraph*{Classical inverse rendering.} %

The earliest technique is {\em shape-from-shading}~\cite{Horn:1989:OSS:93871.93877}, which estimates shape from the shading in an input image with known illumination. 
While methods \revb{in the early years} assume simple illumination models such as point, directional, or area light sources, recent ones adopt environmental illumination represented with \revb{second}-order SH~\cite{DBLP:conf/cvpr/JohnsonA11}.
Also with known shape (e.g., convex shape~\cite{DBLP:conf/iccv/ChandrakerR11}, occluding contour~\cite{DBLP:journals/cgf/Lopez-MorenoGHRG13}, or approximate geometry~\cite{DBLP:journals/tog/KholgadeSES14}), one can estimate reflectance and illumination.
Another mainstream in this literature is {\em intrinsic images}~\cite{Barrow:1978,DBLP:journals/cgf/BonneelKPB17}, which decomposes an input image into shading (i.e., the product of \revd{shape} and illumination) and reflectance based on the Retinex theory~\cite{Land:71}. 
With this \revb{decomposition,} we can change the color or texture while retaining the shading. 
However, for relighting, we must further decompose the shading into shape and illumination.

\paragraph*{Data-driven approaches.} %

Data-driven approaches are commonly adopted in recent techniques for, e.g., outdoor/indoor illumination estimation~\cite{DBLP:conf/cvpr/Hold-GeoffroySH17,DBLP:journals/tog/GardnerSYSGGL17}, estimation of specular \revb{reflectance} and illumination~\cite{DBLP:conf/eccv/OxholmN12,rematas2017reflectancelearning} as well as intrinsic images~\cite{DBLP:journals/tog/BalaS14,DBLP:conf/iccv/NarihiraMY15,DBLP:journals/corr/abs-1712-01056,DBLP:conf/cvpr/ShiDSY17}.
As a generalization of both shape-from-shading and intrinsic images, Barron and Malik~\shortcite{DBLP:journals/pami/BarronM15} factored single input images of general objects into shape, diffuse reflectance, and SH illumination, via optimization with statistical priors.

\paragraph*{Face inverse rendering.} %

\revb{Simultaneous} inference similar to Barron and Malik's work has been actively studied in the inverse rendering of human \revb{faces} since the seminal work of the 3D morphable model (3DMM)~\cite{DBLP:conf/siggraph/BlanzV99}.
The 3DMM is a statistical model of albedo and shape of human \revb{faces} and serves as a strong prior for face inverse rendering via geometric fitting to the target face image.
While the illumination model used in the original 3DMM paper~\cite{DBLP:conf/siggraph/BlanzV99} was directional light, currently the standard choice is again \revb{second}-order SH.
\revd{Due to} the increase of large-scale publicly-available face datasets, many learning-based methods with~\cite{DBLP:conf/iccv/TewariZK0BPT17} and without~\cite{DBLP:conf/cvpr/ShuYHSSS17,DBLP:conf/CVPR2018/abs-1712-01261} 3DMMs have been proposed for face inverse rendering.

Our work also adopts \revb{second}-order SH illumination, but tackles inverse rendering of not only faces but also full bodies including garments.
Full body images contain face regions, and thus existing techniques for faces can be applied to infer illumination.
However, one concern is that most of the existing techniques assume \revd{that} light occlusion \revb{is ignorable}; this assumption might be valid for faces because most faces are approximately convex except for the vicinity of noses, but it \revb{does not hold} true for concave regions in the human body, e.g., armpits, a crotch, or a neck under a chin, that should receive less light due to self-shadowing.
Consequently, such concave regions become unnaturally bright if \revb{we ignore the} light occlusion.
For better relighting, we learn light occlusion for SH-based shading.


Schneider et al.~\shortcite{DBLP:conf/iccv/SchneiderSEFV17} also proposed to account for light occlusion in SH-based face inverse rendering to better handle face wrinkles.
They extended a 3DMM~\cite{DBLP:conf/avss/PaysanKARV09} so that not only albedo and shape but also per-vertex light transport vectors can be reconstructed via multilinear regression.
However, light transport vectors are available only in the face region.


Apart from the SH-based formulation, Yamaguchi et al.~\shortcite{Ymaguchi:SIG18} inferred a base mesh and high-quality textures of \revb{a} face from a single image.
Without considering lighting formulation, they infer textures for photorealistic rendering of faces using regression with \revb{an} adversarial loss. 
Their method relies on \revb{plenty} of high-quality measured data, which are unfortunately not \revb{available} in general for human bodies.

\paragraph*{Other human-oriented techniques.} %

Traditionally, human whole-body relighting has been performed based on measurement under controlled setups with multiple lights and cameras~\cite{DBLP:conf/siggraph/DebevecHTDSS00,DBLP:journals/cgf/LiWSLVDT13}.
In monocular settings, RGB video cameras are also used for capturing faces with multiple temporal frames, e.g., \cite{DBLP:journals/tog/GarridoVWT13}.
Here we focus on single-image techniques.
If the \revb{human target} figure is almost naked, we can obtain a reasonable shape cue for inverse rendering by fitting statistical 3D body models~\cite{DBLP:journals/tog/AnguelovSKTRD05,DBLP:conf/cvpr/BalanSBDH07} after segmenting out the figure mask~\cite{DBLP:conf/iccv/GuanWBB09}.
However, this is generally not applicable to human figures wearing garments.
There are also techniques that can estimate garment shapes from single images~\cite{DBLP:journals/cgf/ZhouCFGT13,DBLP:journals/cgf/DanerekDOZG17}.
Our method is versatile and can capture garment wrinkles \revb{plausibly} from various human portraits.

\paragraph*{CNN-based techniques for material inference.} %

Recent methods can infer materials~\cite{DBLP:journals/tog/AittalaAL16,DBLP:journals/tog/LiDPT17} of objects using CNNs from a single image of flat-surface objects. 
Innamorati et al.~\shortcite{DBLP:journals/cgf/InnamoratiRWM17} proposed an interesting approach that decomposes an input image into multiple components for manual photo retouching.
They account for light occlusion in the form of ambient occlusion and decompose the shading component into six directions based on non-negative \revb{first}-order SH bases.
With this formulation, photo-retouch artists can emulate relighting by manually increasing/decreasing directional shading components.
Inspired by their work, we will compare our method with the conventional SH formulation plus ambient occlusion in Section~\ref{sec:Experiments}. 

\section{Spherical Harmonics (SH) Lighting}
\label{sec:SphericalHarmonicsLighting}

In this section, we briefly review \revb{spherical harmonics} (SH) lighting with and without consideration of light occlusion.

\subsection{SH Lighting without Occlusions}
\label{sec:SHWithoutOcclusions}

SH are orthonormal basis functions defined on \revb{the} spherical domain, and known as advantageous for capturing low-frequency signals in the rendering community.
It is shown that just nine SH bases (i.e., basis functions up to \revb{second} order) can \revd{capture up to 99.22\% of the} irradiance on a convex surface~\cite{1177153}.

Let us review the mathematical formulation~\cite{DBLP:conf/siggraph/RamamoorthiH01a}. 
If we ignore light occlusion and interreflection, the irradiance $E(\mathbf{n})$ can be calculated with an integral of arbitrary incoming radiance $L(\mathbf{\omega}_i)$ over the hemispherical domain $\Omega(\mathbf{n})$ defined by a unit normal vector $\mathbf{n}$
\begin{align}
E(\mathbf{n}) = \int_{\Omega(\mathbf{n})} L(\mathbf{\omega}_i) \, \max(\mathbf{n} \cdot \mathbf{\omega}_i, 0) \, d\mathbf{\omega}_i. 
\label{eq:RenderingEqWoOcclusion}
\end{align}
We omit the dependency on surface position for simplicity.
Ramamoorthi and Hanrahan projected the spherical signals of the incoming illumination distribution $L(\mathbf{\omega}_i)$ and the cosine decay term $\max(\mathbf{n} \cdot \mathbf{\omega}_i, 0)$ to \rev{SH}.
Using elevation and azimuth angles $\theta$, $\phi$ to parameterize a unit direction vector $\mathbf{\omega} = (\theta, \phi)$, these signals are expanded as
\begin{align}
L(\theta, \phi) = \sum_{l,m} L_{l,m} \, Y_{l,m} (\theta, \phi), \\
A(\theta) = \max(\cos \theta, 0) = \sum_l A_l  \, Y_{l,0} (\theta),
\end{align}
where $Y_{l,m}$ are \rev{SH} with $l \ge 0$, $-l \le m \le l$, and $m \le 2$. 
$L_{l,m}$ and $A_l$ are coefficients for the illumination and cosine decay term, respectively.
$A(\theta)$ does not depend on \revd{the} azimuth angle $\phi$. 
The integral in Equation~(\ref{eq:RenderingEqWoOcclusion}) is now rewritten as
\begin{align}
E(\theta, \phi) = \sum_{l,m} \hat{A}_l \, L_{l,m} \, Y_{l,m} (\theta, \phi),
\end{align}
where $\hat{A}_l = \sqrt{\frac{4 \pi}{2 l + 1}} A_l$. 
Here $Y_{l,m}$ can be represented as polynomials of \red{coordinates of a unit normal $\mathbf{n} = (x, y, z)^T$}.
If we rewrite the coefficients \red{$\{ L_{l,m} \}$} as a vector \red{$\mathbf{L}$} and the basis functions \red{$\{ \hat{A}_l  Y_{l,m} \}$} as a vector \red{$\hat{\mathbf{Y}}$}, $E$ is calculated as a dot product
\begin{align}
E = \hat{\mathbf{Y}}^T \mathbf{L}.
\label{eq:DotProdWoOcclusion}
\end{align}
%

\subsection{SH Lighting with Occlusions}
\label{sec:SHWithOcclusions}

Although the above formulation is elegant, the critical problem is that light occlusion is ignored. 
\red{Concave regions should receive less light due to self-shadowing, and thus should be darker than other convex regions.}
To account for light occlusion in Equation~(\ref{eq:RenderingEqWoOcclusion}), the visibility term $V(\mathbf{\omega}_i)$ \red{should be} added in the integrand
\begin{align}
E(\mathbf{n}) = \int_{\Omega(\mathbf{n})} L(\mathbf{\omega}_i) \, V(\mathbf{\omega}_i) \, \max(\mathbf{n} \cdot \mathbf{\omega}_i, 0) \, d\mathbf{\omega}_i. 
\label{eq:RenderingEqWithOcclusion}
\end{align}
$V(\mathbf{\omega}_i)$ returns \revb{zero} if \revb{the} light in the incoming direction $\mathbf{\omega}_i$ is occluded and \revb{one} otherwise.
Unfortunately, $V(\mathbf{\omega}_i)$ does not have any analytical form in general, and one must sample visibility by casting many shadow rays at each surface point, which is quite computationally expensive.

Sloan et al.~\shortcite{DBLP:journals/tog/SloanKS02} proposed to precompute the visibility term together with the cosine decay term, and project the compound spherical signal onto \rev{SH} in order to enable efficient dot-product calculation (similar to Equation~(\ref{eq:DotProdWoOcclusion})) during real-time rendering
\begin{align}
E = \mathbf{T}^T \mathbf{L},
\label{eq:DotProdWithOcclusion}
\end{align}
where $\mathbf{T}$ is a vector that encodes SH coefficients of the compound spherical signal of the visibility term and the cosine decay term.
They also proposed to handle glossy reflection and approximate interreflection.
This \revb{technique}  is \revd{well-known as} {\em precomputed radiance transfer} (PRT), which \revd{has} been studied and extended extensively in the real-time rendering literature.

\red{Hereafter we refer to $\mathbf{T}$ as a {\em light transport vector} and a nine-channel image containing per-pixel light transport vectors as a {\em light transport map}.}

\section{Our Loss Functions}
\label{sec:OurLossFunctions}

\begin{figure*}[t]
  \centering
  \includegraphics[width=0.9 \linewidth]{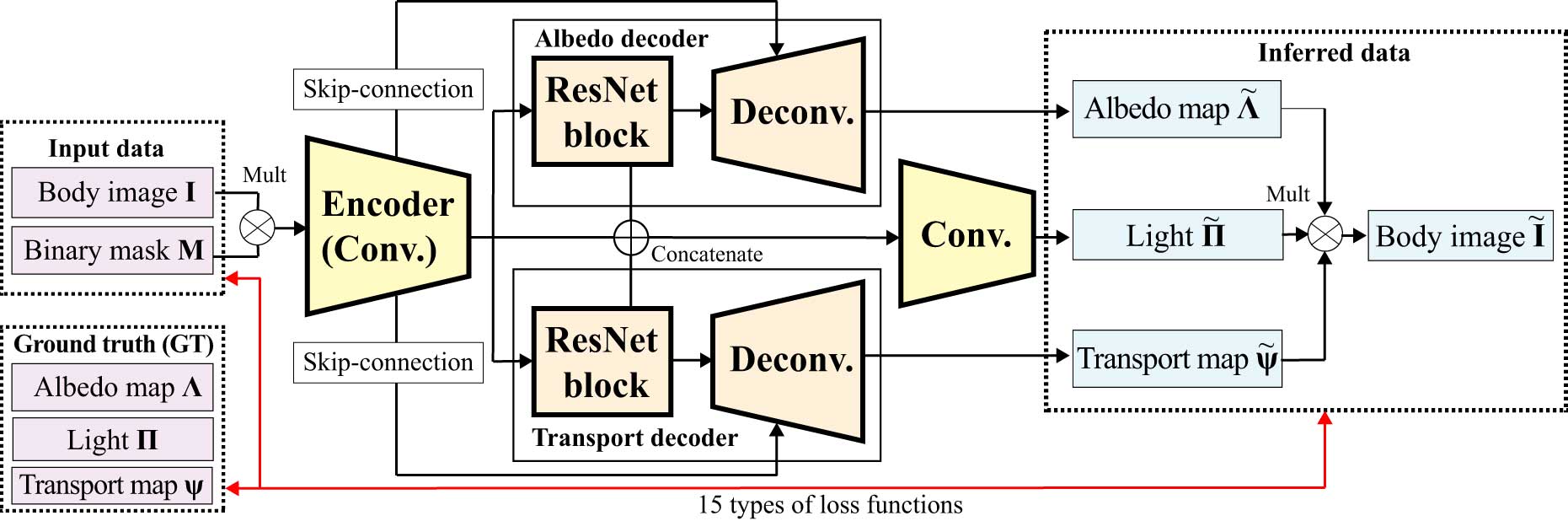}
  \caption{Our network architecture. \revb{The input image (multiplied with the binary mask) is fed to the encoder, and the output feature vector is then fed to the decoders of albedo, light transport, and light (Section~\ref{sec:NetworkModels}). We calculate 15 types of loss functions by considering the total variations (TVs) for albedo and light transport maps as well as the combinations of inferred outputs and their ground-truth (GT) (Section~\ref{sec:OurLossFunctions}).}}
  \label{fig:NetworkModels}
 \end{figure*}

\red{In this section,} we define the loss functions to infer albedo and light transport maps using our CNNs \red{based on the SH formulation.}

\red{For training and testing, we prepare a synthetic human image dataset $\mathcal{D}_H$ and an illumination dataset $\mathcal{D}_L$ (see Section~\ref{sec:DatasetGeneration} for the details).
The synthetic human image dataset $\mathcal{D}_H$ contains a set of a binary mask $\mathbf{M}^c_j  \in \{0, 1\}^{N \times c}$ (where $N$ is the number of pixels, $c$ is the number of channels, and $j = 1, 2, \dots, |\mathcal{D}_H|$), albedo map $\mathbf{\Lambda}_j \in \mathbb{R}^{N \times 3}$, and light transport map $\mathbf{\Psi}_j \in \mathbb{R}^{N \times 9}$ \rev{for each 3D human model}.
The illumination dataset $\mathcal{D}_L$ contains SH illumination coefficients for RGB channels $\mathbf{\Pi}_k \in \mathbb{R}^{9 \times 3}$, where $k = 1, 2, \dots, |\mathcal{D}_L|$.
Note that we multiply the binary mask $\mathbf{M}^c_j$ to the ground-truth data and network outputs (e.g., $\mathbf{M}^3_j * \mathbf{\Lambda}_j$ or $\mathbf{M}^9_j * \mathbf{\Psi}_j$, where $*$ denotes element-wise multiplication) so that we can ignore out-of-mask pixels. 
In the following explanation, we omit the element-wise multiplication of the binary mask for simplicity.}

We use a CNN \revb{architecture} for inferring light, albedo, and light transport maps (see Section~\ref{sec:NetworkModels} for the network models). 
The input of the CNN is a masked, RGB full-body human image $\mathbf{I}_{j,k} = \mathbf{\Lambda}_j * ( \mathbf{\Psi}_j \, \mathbf{\Pi}_k )$.  
Let $\tilde{\mathbf{\Lambda}}_{j,k} = \tilde{\mathbf{\Lambda}}(\mathbf{I}_{j,k}; \Theta_{\lambda}) \in \mathbb{R}^{N\times3}$ be the CNN output for an albedo map, $\tilde{\mathbf{\Psi}}_{j,k} = \tilde{\mathbf{\Psi}}(\mathbf{I}_{j,k}; \Theta_{\psi}) \in \mathbb{R}^{N\times9}$ the CNN output for a light transport map, and $\tilde{\mathbf{\Pi}}_{j,k} = \tilde{\mathbf{\Pi}}(\mathbf{I}_{j,k}; \Theta_{\pi}) \in \mathbb{R}^{N\times9}$ the CNN output for illumination.
\rev{Regarding notations, we use tildes~($\,\tilde{}\,$) to indicate inferred outputs, and denote $\tilde{f}(x;y)$ to indicate that $x$ is the input and $y$ is the parameter of network $\tilde{f}$.}
We optimize these network parameters $\Theta_{\lambda}$, $\Theta_{\psi}$, and $\Theta_{\pi}$ via regression.

Our CNN architecture has a similar design to {\em SfSNet}~\cite{DBLP:conf/CVPR2018/abs-1712-01261}, which infers light, albedo, and normal maps for faces simultaneously.
The loss functions used in SfSNet are \rev{L1} losses for the inferred albedo map, normal map (from which a light transport map without light occlusion can be calculated analytically), light, and the reconstructed image using the three components. 
We also use similar four loss functions, but \rev{we do not infer normal maps but infer} light transport maps \revd{directly}.
Namely, we use L1 losses for $\tilde{\mathbf{\Lambda}}_{j,k}$, $\tilde{\mathbf{\Psi}}_{j,k}$, $\tilde{\mathbf{\Pi}}_{j,k}$, and the reconstructed image $\tilde{\mathbf{I}}_{j,k} = \tilde{\mathbf{\Lambda}}_{j,k} * ( \tilde{\mathbf{\Psi}}_{j,k} \, \tilde{\mathbf{\Pi}}_{j,k} )$.
Furthermore, we also use the following L1 losses:
\begin{description}
\item[\rev{TV losses:}] L1 total variation \rev{(TV) losses} both for albedo $\tilde{\mathbf{\Lambda}}_{j,k}$ and light transport maps $\tilde{\mathbf{\Psi}}_{j,k}$,

\item[\rev{Shading losses:}] Three patterns of combination of inferred/GT data to compute a shading map, i.e., 
$\tilde{\mathbf{\Psi}}_{j,k} \, \mathbf{\Pi}_k$, 
$\mathbf{\Psi}_j \, \tilde{\mathbf{\Pi}}_{j,k}$, and 
$\tilde{\mathbf{\Psi}}_{j,k} \, \tilde{\mathbf{\Pi}}_{j,k}$, 

\item[\rev{Reconstruction losses:}] Six patterns of combination to reconstruct an input image, i.e., 
$\mathbf{\Lambda}_j * ( \mathbf{\Psi}_j \, \tilde{\mathbf{\Pi}}_{j,k} )$,
$\mathbf{\Lambda}_j * ( \tilde{\mathbf{\Psi}}_{j,k} \, \mathbf{\Pi}_k )$,
$\mathbf{\Lambda}_j * ( \tilde{\mathbf{\Psi}}_{j,k} \, \tilde{\mathbf{\Pi}}_{j,k} )$,
$\tilde{\mathbf{\Lambda}}_{j,k} * ( \mathbf{\Psi}_j \, \mathbf{\Pi}_k )$,
$\tilde{\mathbf{\Lambda}}_{j,k} * ( \mathbf{\Psi}_j \, \tilde{\mathbf{\Pi}}_{j,k} )$, and
$\tilde{\mathbf{\Lambda}}_{j,k} * ( \tilde{\mathbf{\Psi}}_{j,k} \, \mathbf{\Pi}_k )$.
\end{description}
In total, we use 15 L1 losses.
All weights are set to one.

To consider the benefit of the 15 losses, let us take the shading losses, i.e., the three losses for a shading map, as an example. 
For the multiplication of a light transport map and a light, there are three combinations, namely, GT * inferred (i.e., $\mathbf{\Psi}_j \, \tilde{\mathbf{\Pi}}_{j,k}$), inferred * GT (i.e., $\tilde{\mathbf{\Psi}}_{j,k} \, \mathbf{\Pi}_k$), and inferred * inferred (i.e., $\tilde{\mathbf{\Psi}}_{j,k} \, \tilde{\mathbf{\Pi}}_{j,k}$), where GT means ground truth. 
If GT is involved, GT works as a weighting matrix for the inferred output, which enforces the output to lie on a solution manifold in the high-dimensional space. 
If both are inferred outputs, the loss becomes a soft constraint for the intermediate output, i.e., the inferred shading map. 
For the choice of loss functions, we chose formulae that do not introduce bias, except for the TV losses.
\revb{In this way, involving} as many formulae \revd{that do not introduce bias} as possible as losses is \revb{a} quite general \revb{technique}, and would be beneficial to other problems.
We show an ablation study with and without these losses in Section~\ref{sec:Comparisons}.

\section{Network Models}
\label{sec:NetworkModels}



Figure~\ref{fig:NetworkModels} illustrates our encoder-decoder network.
As mentioned, our network is similar to that of SfSNet, except that ours has much \revd{more} parameters.
Our encoder has six convolutional layers whose output channels are \{ 64, 128, 256, 512, 512, 512 \} and the stride is two.
The encoded features are then fed to the decoders for albedo, light transport, and light.
The decoders for albedo and light transport maps have almost the same architecture, except that the numbers of output channels are different (i.e., nine for light transport and three for albedo). 
Each decoder has a residual block (consisting of two convolutional layers with 512 channels) and six deconvolutional layers (output channels are \{ 512, 512, 256, 128, 64, 9 or 3 \} and the stride is also two).
The encoder and decoders are connected using skip-connections.
For the light decoder, the outputs of the encoder and decoders for albedo and light transport are concatenated and fed to four convolutional layers, which yield a 27-dimensional vector.
While SfSNet uses average pooling layers, ours consists of (de-)convolutional layers only.
Each (de-)convolutional layer (except for the first and final layers) is followed by batch normalization and (leaky) ReLU.
The first three deconvolutional layers of each decoder are followed by dropout with probability 0.5.

%

\section{Dataset Generation}
\label{sec:DatasetGeneration}

\red{As explained in Section~\ref{sec:OurLossFunctions}, we prepared a synthetic human image dataset and an illumination dataset.
Here we explain the details.}

\begin{figure}[t]
  \centering
  \includegraphics[width=\linewidth]{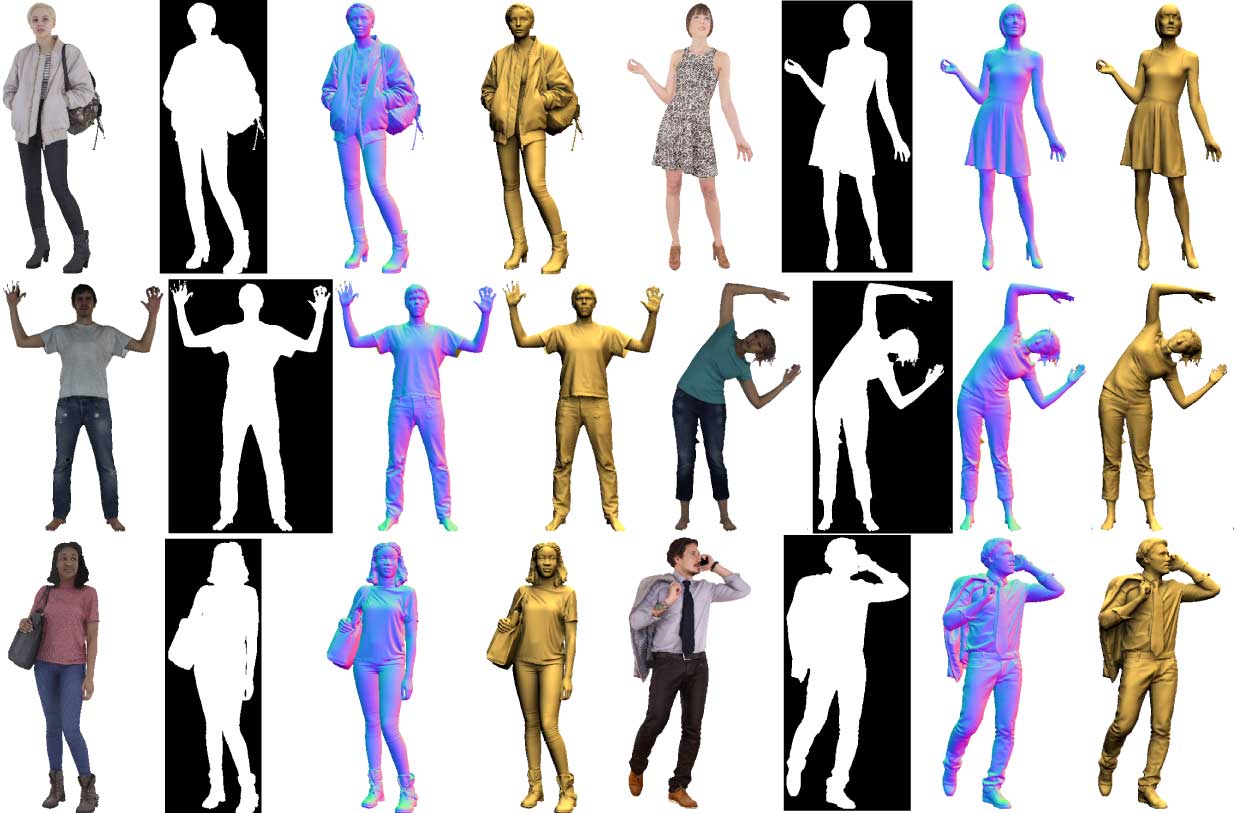}
  \caption{Examples from our synthetic human image dataset. For \revd{each} human figure, an albedo map, binary mask, normal map, and light transport map (visualized as a shading map) are displayed. \revb{Note that some albedo maps contain self-shadows due to scanning inadequacy.} Each image is trimmed. }
  \label{fig:HumanImageDataset}
 \end{figure}

\paragraph{Synthetic human image dataset.} %
\red{Our synthetic human image dataset consists} of a binary mask, albedo map, normal map, and light transport map, created by rendering each scanned 3D human figure \red{using a hardware-accelerated renderer}.
The scanned 3D human figures were obtained from two resources; one is the publicly-available {\em BUFF dataset}~\cite{DBLP:conf/cvpr/ZhangPBP17}, and the other is commercial websites.
The BUFF dataset contains \red{9,613 standing} 3D figures, but lacks variations for our purpose.
Namely, it only includes \red{five} individuals with one or two \revd{outfits each} and time-varying poses, and thus subsequent 3D models of the same individual in the same \revd{outfit} are almost identical.
To avoid biasing the training dataset, we manually picked \revd{74 representative} models from the BUFF dataset.
The commercial data were purchased from different \revb{websites} and amount to \red{271} models.
We randomly split \revd{the models, 345 in total,} into 276 training data and 69 test data.
Figure~\ref{fig:HumanImageDataset} shows some examples of our training data.
Note that some albedo maps contain \revb{self-shadows} because shading was not completely removed during the scanning process. 

When creating the dataset, we carefully aligned 3D models so that our CNNs can exploit the geometric regularity of our small dataset.
\rev{Namely, we rendered front-facing figures in the middle of square images while aligning them so that they have almost the same vertical size in the images with vertical paddings at the top and bottom of a fixed size (5\% of image heights).}
Regarding poses, we only used standing figures and removed sitting ones \rev{from our training/test datasets}. 
The image resolution is \rev{$1024 \times 1024$ pixels.}
\rev{No data augmentation is employed for the human image dataset.}

\begin{figure}[t]
  \centering
  \includegraphics[width=\linewidth]{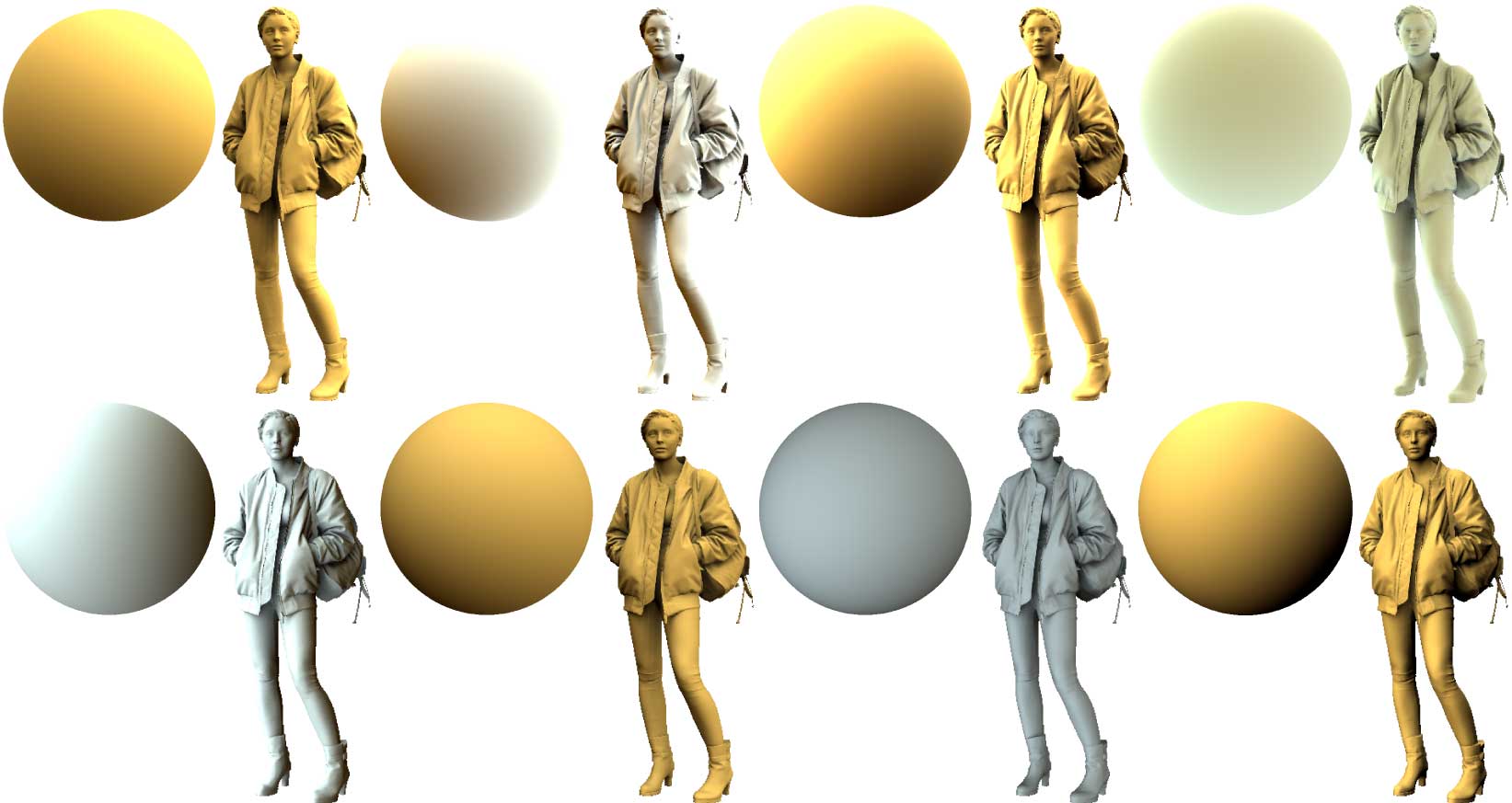}
  \caption{Examples from our illumination dataset. SH illumination coefficients are visualized as shading maps of a sphere and the top-left model in Figure~\ref{fig:HumanImageDataset}.}
  \label{fig:IlluminationDataset}
 \end{figure}

\paragraph{Illumination dataset.} %
For \red{our} illumination dataset, we used the \red{{\em Laval Indoor HDR dataset}}~\cite{DBLP:journals/tog/GardnerSYSGGL17} containing 2,144 environment maps in panoramic HDR format.
We first converted them into diffuse SH \revb{coefficients} and calculated a reference brightness of each environment map using Equation~(\ref{eq:DotProdWoOcclusion}) with a front-facing normal $\mathbf{n} = (0,0,1)^T$.
We omitted dark environment maps if the reference brightness is lower than 0.2, and scaled the brightness of other environment maps so that reference brightness lies within $[0.7, 0.9]$.
\red{To obtain further variations, we rotated each data 35 times by 10 degrees around the vertical axis.
We then reduced the redundancy using \revb{k-means clustering} and manually removed unusual illuminations (e.g., too \revb{bright} lights, \revb{back-lights}, \revb{and} lights causing too strong contrasts in shadings).
Finally, from \revb{the} remaining 50 illuminations, we randomly picked 40 illuminations for training and 10 for testing.
Figure~\ref{fig:IlluminationDataset} shows some examples of our training data.}

\section{Experiments}
\label{sec:Experiments}

\begin{table*}[]
\centering
\caption{RMSE and SSIM for the inferred results using each method. The light transport maps of SfSNet and SfSNet-AO were computed from corresponding normal maps analytically. \rev{Best values are highlighted in boldface.}}
\label{fig:RMSEandSSIM}
\scalebox{0.92}[0.92]{
\begin{tabular}{l||cccccc|cccccc}
 & \multicolumn{6}{|c|}{RMSE within binary masks} & \multicolumn{6}{c}{SSIM within bounding boxes of masks} \\ \cline{2-13}
 & Shading & Transport & Normal & AO & Light & Albedo & Shading & Transport & Normal & AO & Light & Albedo \\ \hline \hline
SfSNet & 0.299 & 0.526 & {\bf 0.346} & N/A & 0.207 & 0.135 & 0.884 & 0.755 & {\bf 0.776} & N/A & 0.446 & 0.954\\\hline
SfSNet-AO & 0.293 & 0.529 & 0.347 & {\bf 0.083} & 0.207 & 0.131 & 0.890 & 0.749 & 0.772 & {\bf 0.946} & 0.475 & \rev{\bf 0.955}\\ \hline
Ours (min) & 0.237 & 0.406 & N/A & N/A & 0.205 & 0.131 & 0.909 & 0.777 & N/A & N/A & 0.473 & 0.953\\ \hline
Ours (full) & {\bf 0.219} & {\bf 0.393} & N/A & N/A & {\bf 0.199} & {\bf 0.129} & {\bf 0.927} & {\bf 0.781} & N/A & N/A & {\bf 0.500} & 0.943 \\ \hline
\end{tabular}
}
\end{table*}

\begin{table*}[]
\centering
\caption{\rev{RMSE and SSIM for an ablation study for our 15-losses. Best values are highlighted in boldface.}}
\label{tab:AblationStudy}
\scalebox{0.92}[0.92]{
\begin{tabular}{l||cccc|cccc}
 & \multicolumn{4}{|c|}{RMSE} & \multicolumn{4}{c}{SSIM} \\ \cline{2-9}
 & Shading & Transport & Light & Albedo & Shading & Transport & Light & Albedo \\ \hline \hline
W/o TV & 0.226 & {\bf 0.391} & 0.202 & {\bf 0.126} & 0.923 & {\bf 0.784} & 0.471 & {\bf 0.956} \\ \hline
W/o shading & 0.227 & 0.398 & 0.201 & 0.132 & 0.922 & 0.781 & 0.496 & 0.940 \\ \hline
W/o reconstruction & {\bf 0.224} & 0.394 & {\bf 0.198} & 0.144 & {\bf 0.925} & 0.782 & {\bf 0.503} & 0.907 \\ \hline
\end{tabular}
}
\end{table*}



\begin{figure*}[t]
  \centering
  \includegraphics[width=\linewidth]{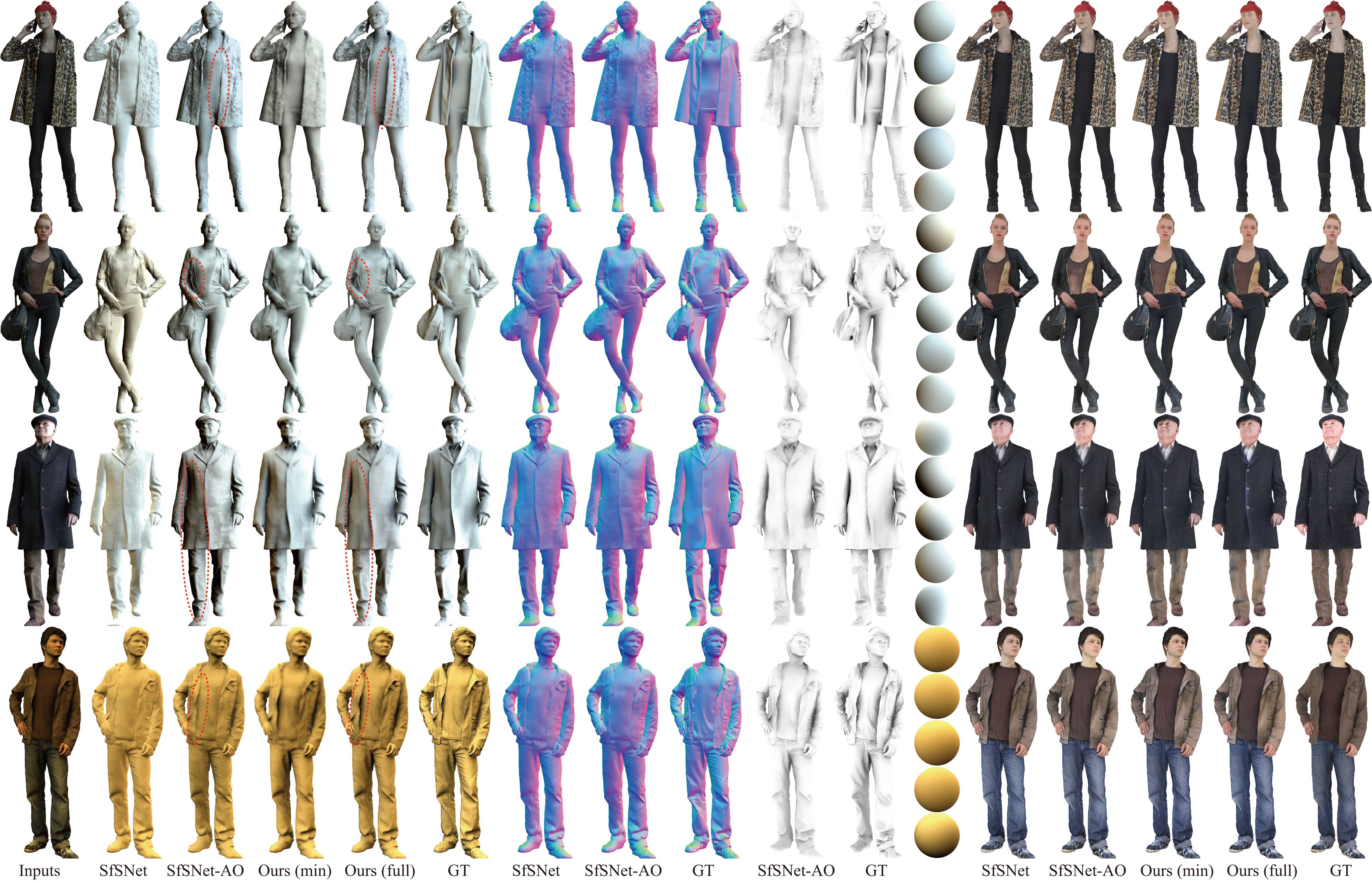}
  \caption{Comparisons with synthetic data. From left to right, input images, shading maps, normal maps, ambient occlusion maps, light maps (i.e., light information visualized by shading a sphere), and albedo maps. The light maps are in the same order as shading/albedo maps, from top to bottom. \revb{The red ovals in inferred shading maps highlight differences between SfSNet-AO and ``Ours (full).''}}
  \label{fig:SyntheticResults}
 \end{figure*}

\begin{figure*}[t]
  \centering
  \includegraphics[width=\linewidth]{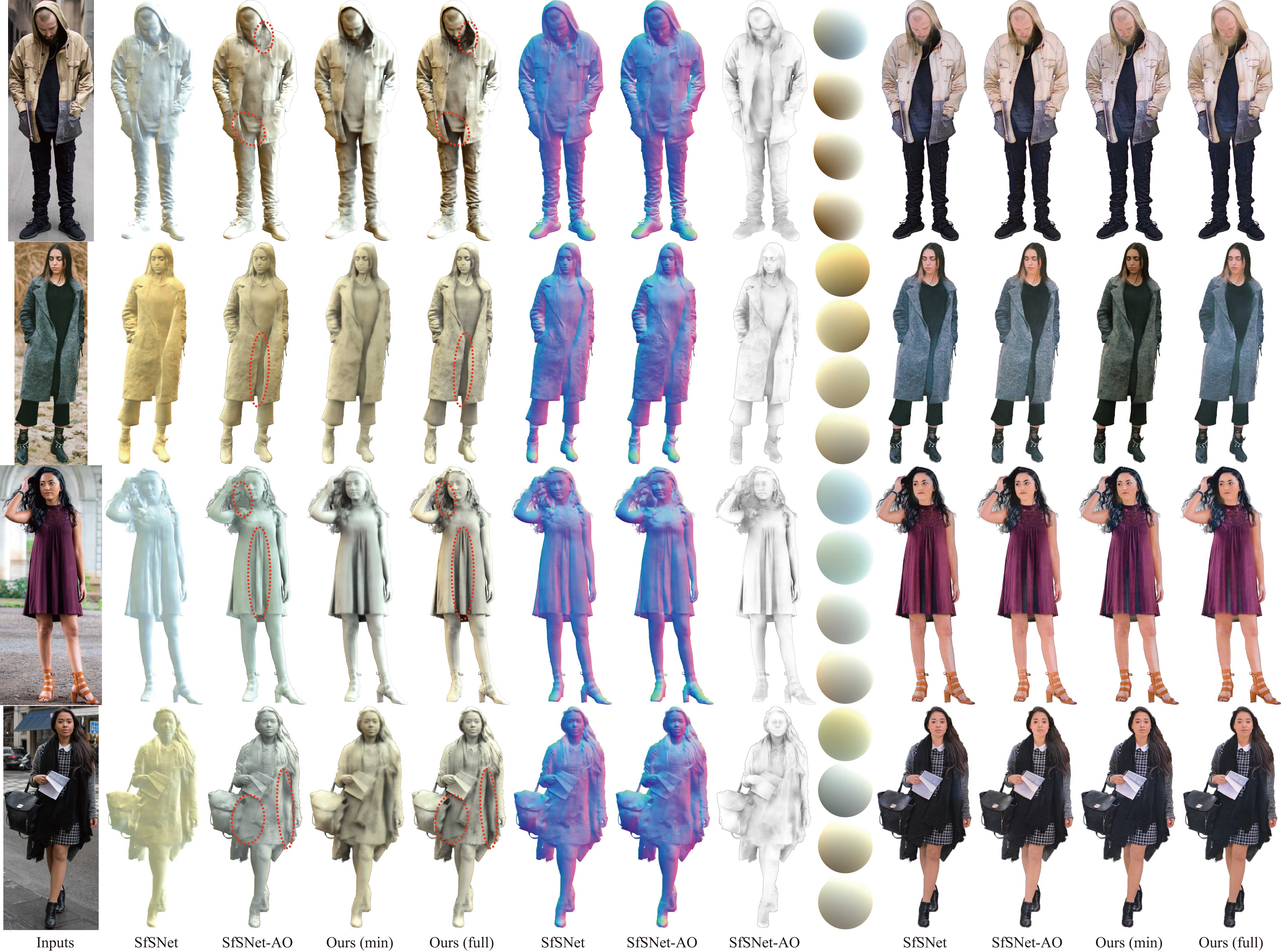}
  \caption{Comparisons with real photographs. \revb{The red ovals in inferred shading maps highlight differences between SfSNet-AO and ``Ours (full).''} Input images courtesy of Guillaume Bolduc, George Gvasalia, Jacob Postuma, and Kat Garcia.}
  \label{fig:ResultsWithPhotographs}
 \end{figure*}


 \begin{figure*}[t]
  \centering
  \includegraphics[width=0.7 \linewidth]{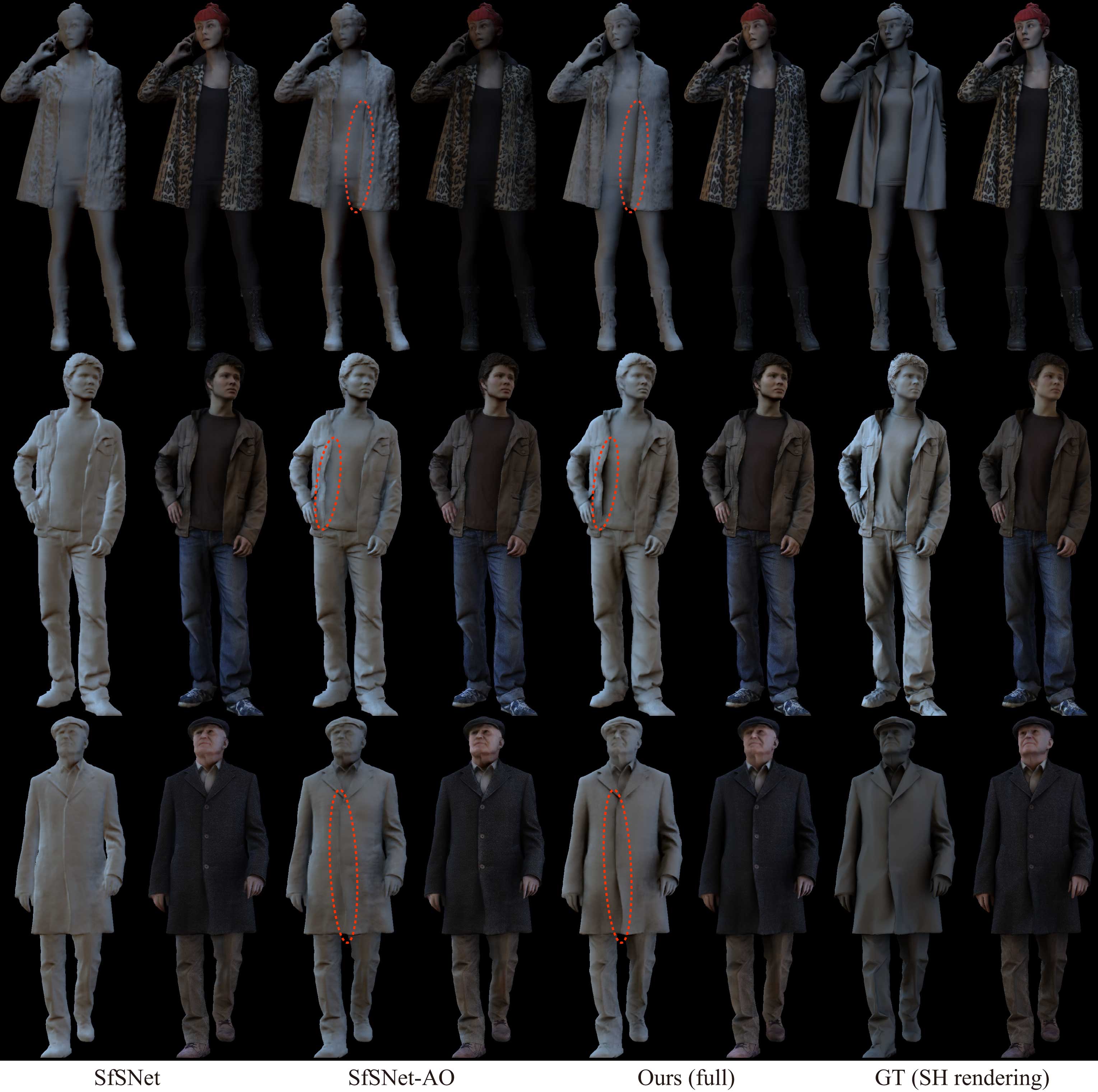}
  \caption{Relighting comparison using synthetic data. \revb{The red ovals highlight differences between SfSNet-AO and ``Ours (full).''}}
  \label{fig:RelightingSynthetic}
 \end{figure*}

 \begin{figure}[t]
  \centering
  \includegraphics[width=\linewidth]{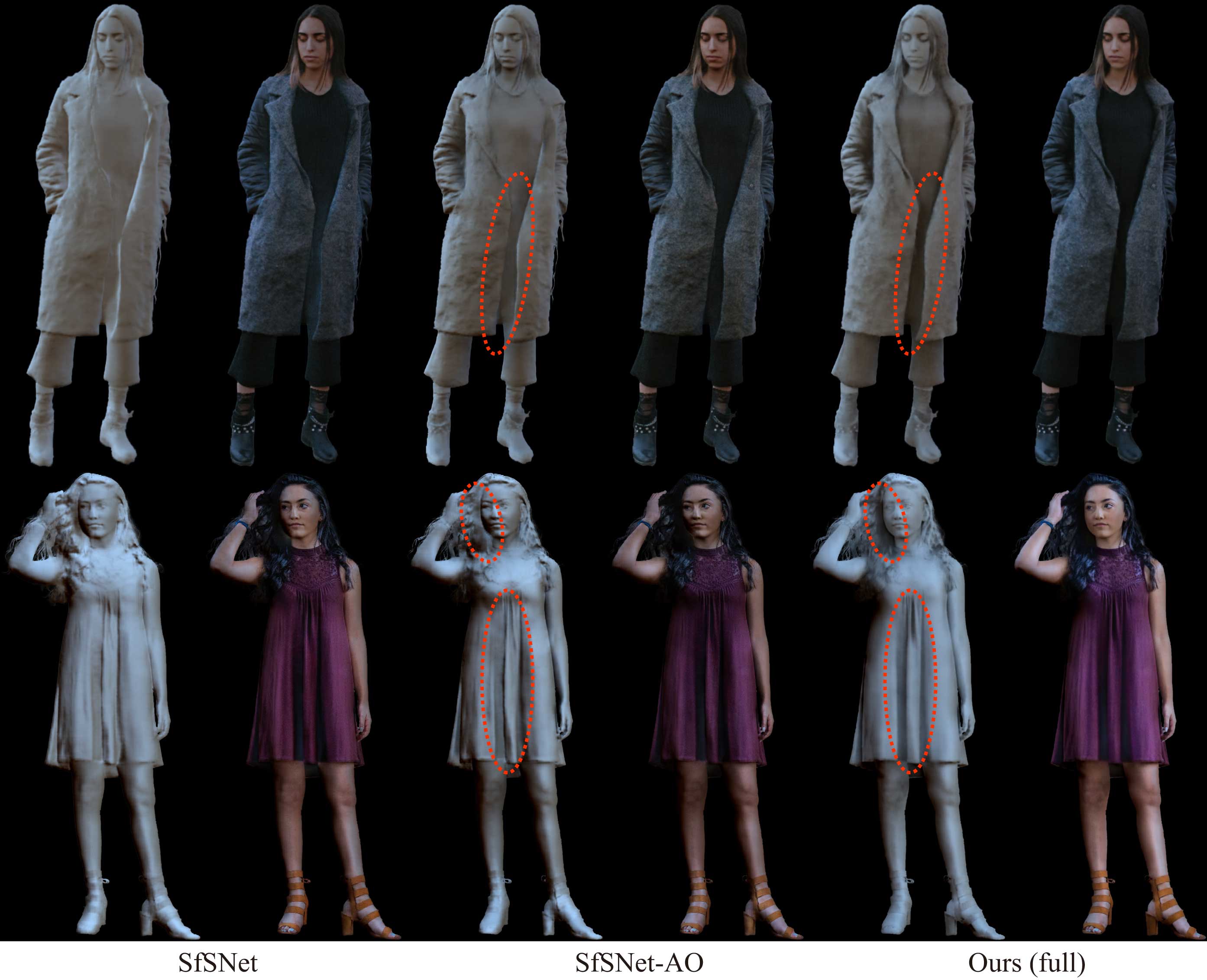}
  \caption{Relighting comparison using real photographs. \revb{The red ovals highlight differences between SfSNet-AO and ``Ours (full).''} Input images courtesy of George Gvasalia and Jacob Postuma.}
  \label{fig:RelightingRealPhoto}
 \end{figure}

We implemented our CNN models using Python and the chainer library, and ran our code on a PC with NVIDIA GeForce GTX 1080 Ti GPUs.
We used Adam as an optimizer with a fixed learning rate 0.0002 and batch size 1.
The computation times for one epoch of training on a single GPU were about three hours with our CNN models.
We used the synthetic images of $1024 \times 1024$ pixels for training in our results.
Our CNN models, as well as other models for comparisons, were trained up to 60 epochs. 
For relighting, we used \revb{Debevec's environment maps~\shortcite{LightProbe2004}}, namely, \texttt{kitchen\_probe} for Figures~\ref{fig:Teaser}, \ref{fig:RelightingSynthetic}, and \ref{fig:RelightingRealPhoto} and \texttt{grace\_probe} for Figure~\ref{fig:Teaser}.
The input photographs in our results were downloaded from {\em Unsplash}\footnote{\url{https://unsplash.com/}}.
Specifically, we selected high-quality free-license images of single human figures, generated their binary masks automatically using Adobe Photoshop with manual correction, applied trimming and uniform scaling, and then added paddings to make them $1024 \times 1024$ pixels.

\subsection{Comparisons of Inference}
\label{sec:Comparisons}

To clarify the advantage of our method, we compared it with three alternative methods.
The first one is {\em SfSNet}~\cite{DBLP:conf/CVPR2018/abs-1712-01261}, but the network architecture is not the original one for small \rev{$128 \times 128$} images but much richer one defined in Section~\ref{sec:NetworkModels}.
In this case, a decoder of SfSNet outputs three-channel normal maps, instead of nine-channel light transport maps.
The second method is SfSNet plus ambient occlusion (hereafter we call it {\em SfSNet-AO}).
A single-channel ambient occlusion is inferred by an additional decoder branch.
The third method is our network with four losses only, similar to SfSNet.
We refer to the 4-loss version as ``Ours (min)'' and the 15-loss version as ``Ours (full).''
Comparisons between ``SfSNet'' and ``Ours (min)'' reveal the impact of considering light occlusion whereas those between ``Ours (min)'' and ``Ours (full)'' demonstrate the effectiveness of \revb{the full loss}.

Figures~\ref{fig:SyntheticResults} and \ref{fig:ResultsWithPhotographs} show the results of qualitative comparisons using synthetic test data and photographs, respectively.
The red ovals in inferred shading maps \revb{highlight} differences between SfSNet-AO and ``Ours (full).''
The first row of Figure~\ref{fig:SyntheticResults} indicates that all methods suffer from separating textures from shading maps.
The shading maps of SfSNet often seem like flat bas-reliefs because light occlusion is ignored. 
In the first and fourth rows, SfSNet-AO estimates the depth gaps between jackets and shirts smaller than the actual gap.
Such biased estimate in shading maps often \rev{yields} unnaturally-darkened albedo maps.
Comparing our two variants, i.e., ``Ours (min)'' and ``Ours (full)'', the latter yields sharper shading maps than the former.
Also in Figure~\ref{fig:ResultsWithPhotographs}, we can see the similar tendency with real photographs.

For quantitative comparison, Table~\ref{fig:RMSEandSSIM} summarizes the RMSE and SSIM of each component.
To reduce the effects of out-of-mask-pixels, \revb{we calculate} RMSEs within binary masks whereas \revb{we calculate} SSIMs within the bounding boxes of binary masks.
The table shows that ``Ours (full)'' is consistently better than other alternatives \rev{except for ``Albedo SSIM''}.
\rev{The reason why ``Albedo SSIM'' of ``Ours (full)'' is lower than others is that ``Ours(full)'' better \revd{cancels the} baked-in shadings (see Section~\ref{sec:DatasetGeneration}) in GT albedos and thus its output albedos become more dissimilar to ``GT.''}
Table~\ref{tab:AblationStudy} further reveals the impacts of the TV losses, shading losses, and reconstruction losses. 
We can see the tendency that overall the accuracies are lower than those of ``Ours (full)'' in Table 1. 
Note that light transport and albedo maps of ``W/o TV'' are slightly better than those of ``Ours (full).'' 
This \revb{result} is \revb{reasonable} because TV losses enforce smoothing, i.e., add biases, to the inferred outputs in compensation for generalization capability.



\subsection{Relighting and Light Transfer}

Figures~\ref{fig:RelightingSynthetic}~and~\ref{fig:RelightingRealPhoto} show the results of relighting with inferred albedo and light transport maps, \revc{given synthetic test images and real photographs, respectively}.
Comparisons with path-traced reference images as well as movies \revd{are} available in the supplemental material.


\begin{figure}[t]
  \centering
  \includegraphics[width=\linewidth]{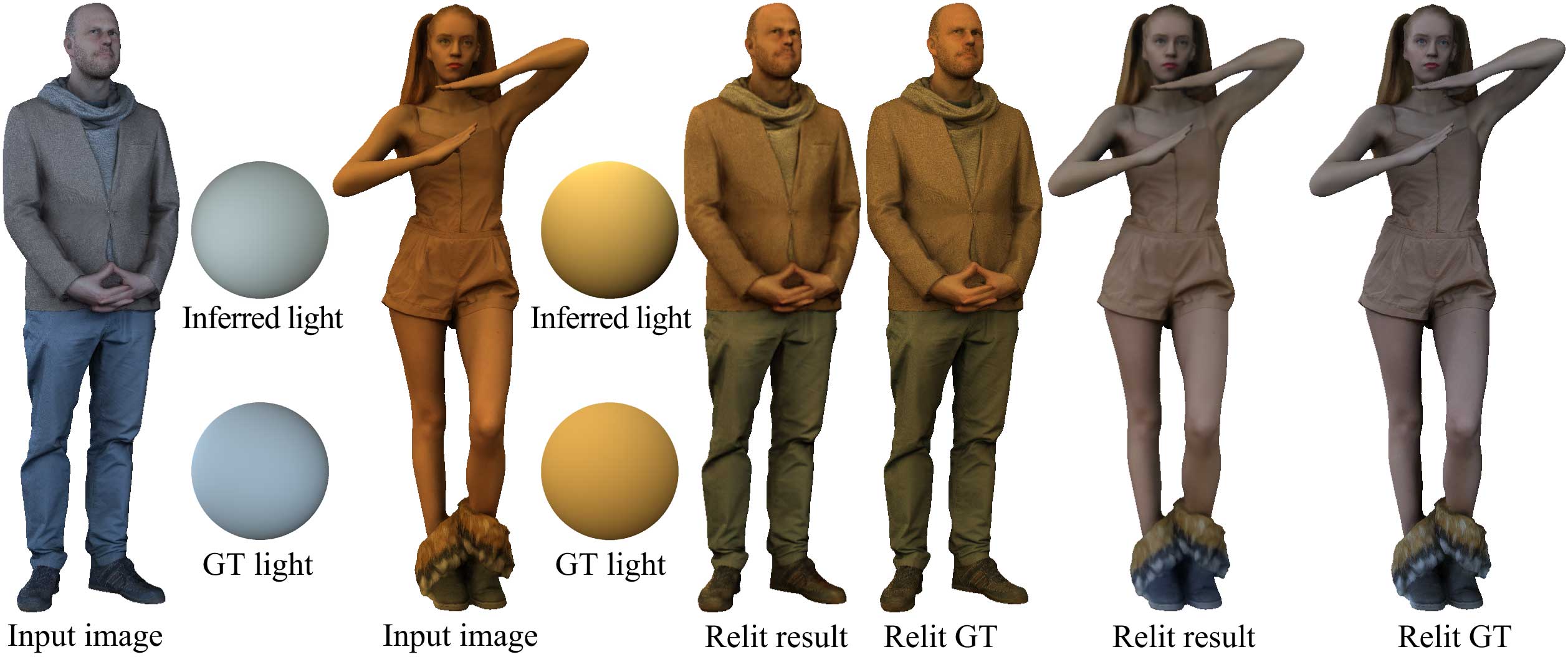}
  \caption{Light transfer. \revb{By performing inference for two images and swapping inferred lights, we can transfer the inferred light to each other.}}
  \label{fig:LightTransfer}
 \end{figure}

\green{By inferring illuminations in two human portraits, we can transfer the inferred illumination to each other.
Figure~\ref{fig:LightTransfer} shows the results of light transfer with synthetic human images.
The inferred illuminations have colors slightly different from the ground-truth, but the patterns of the illuminations are similar.
The relit human images are therefore similar to the ground-truth.
}

\begin{figure}[t]
  \centering
  \includegraphics[width=\linewidth]{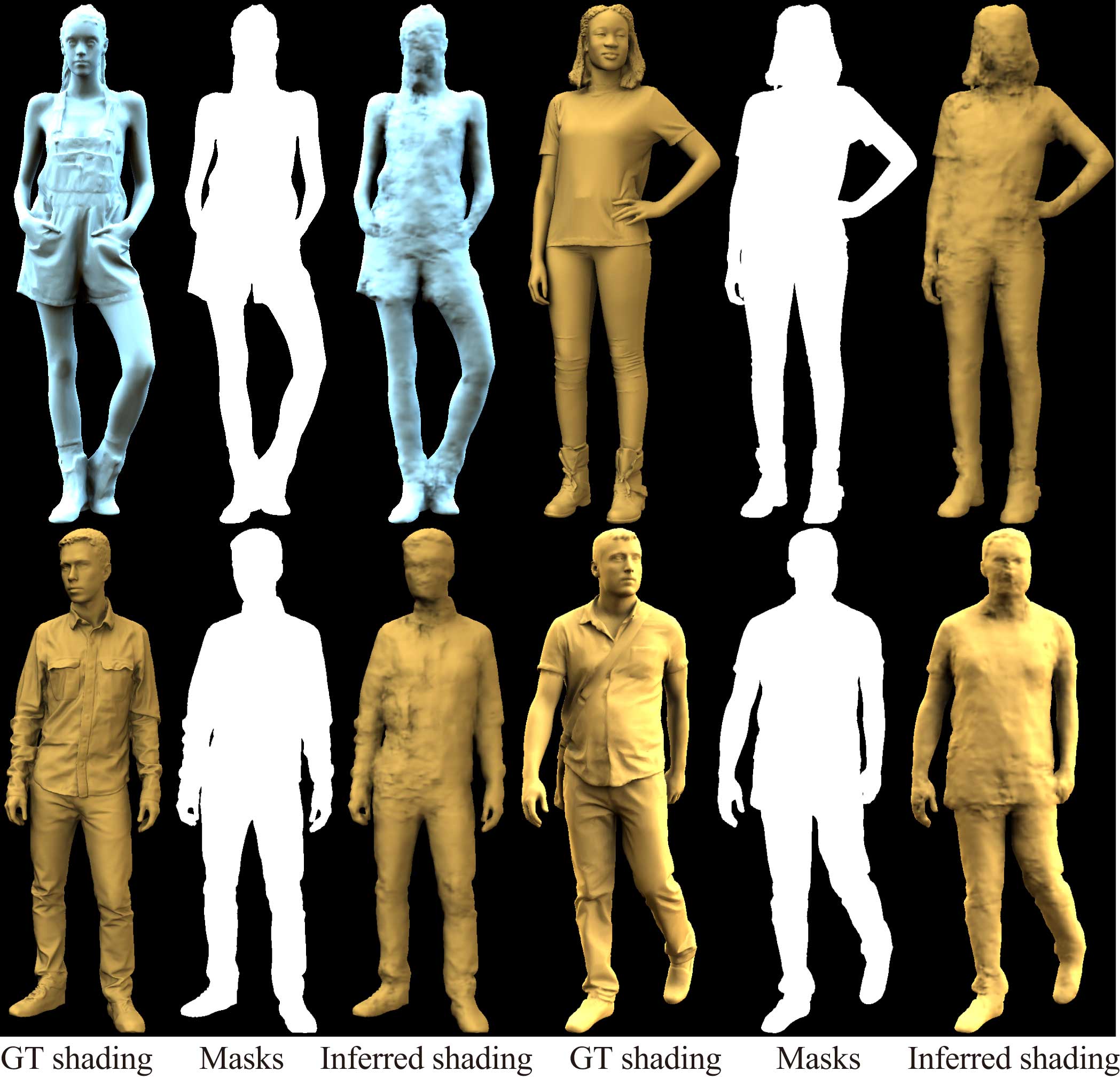}
  \caption{Shape from mask. For each \revb{human figure}, \revb{a} ground-truth shading map, mask, and inferred shading map are displayed. \revb{These results imply that our network can learn a strong shape prior from silhouettes.}}
  \label{fig:ShapeFromMask}
 \end{figure}

\section{Discussions}
\label{sec:Discussions}

\begin{figure}[t]
  \centering
  \includegraphics[width=\linewidth]{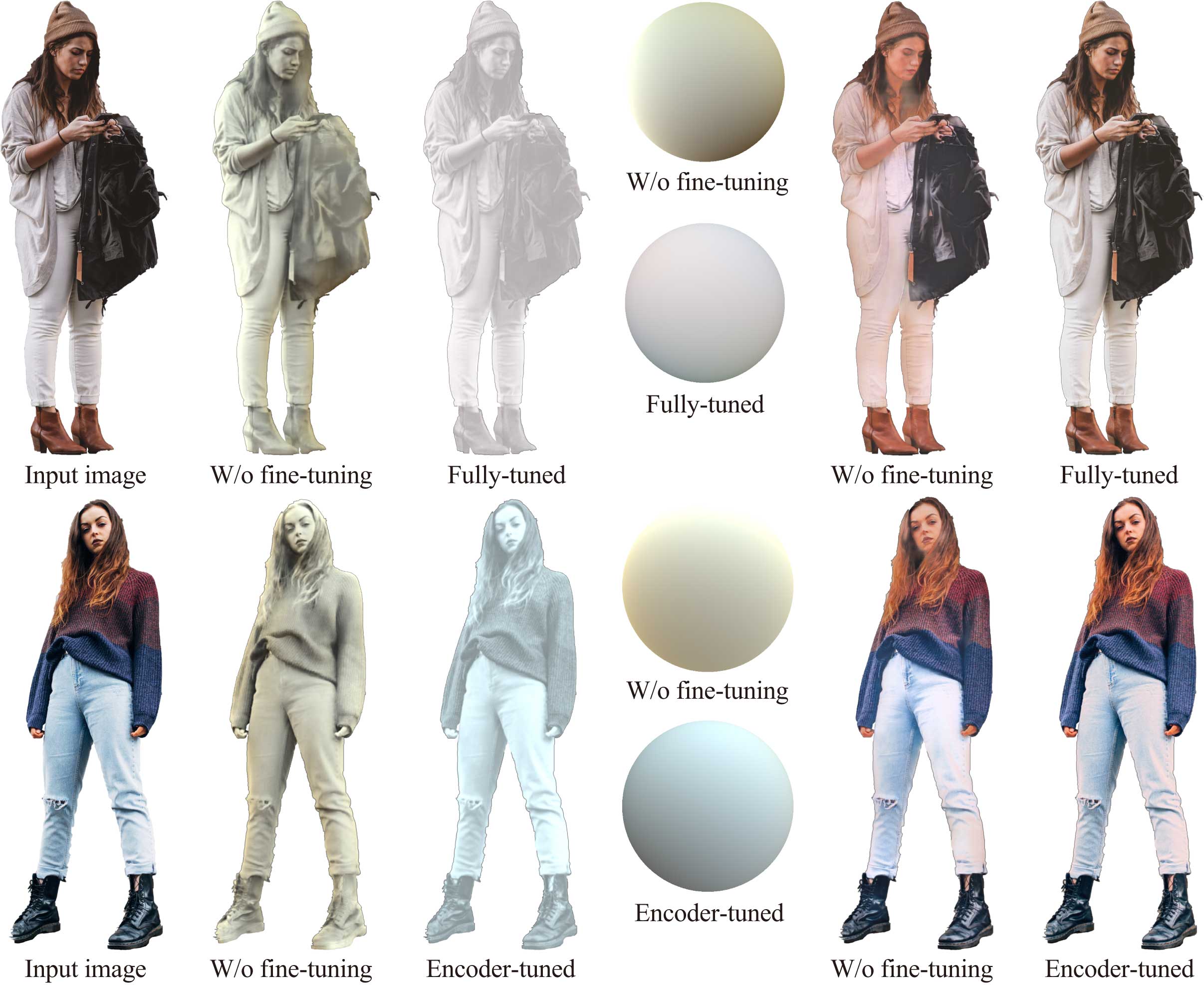}
  \caption{\revb{Comparison to} self-supervised learning. \revb{By fine-tuning network parameters using only unlabeled photographs  with (i.e., encoder-tuned; lower row) and without (i.e., fully-tuned; upper row) fixing decoders, the outputs collapsed; the shading maps bleached and the albedo maps got close to the input images.} Input images courtesy of Philip Martin and Ali Morshedlou.}
  \label{fig:SelfSupervisedLearning}
 \end{figure}

\begin{figure}[t]
  \centering
  \includegraphics[width=\linewidth]{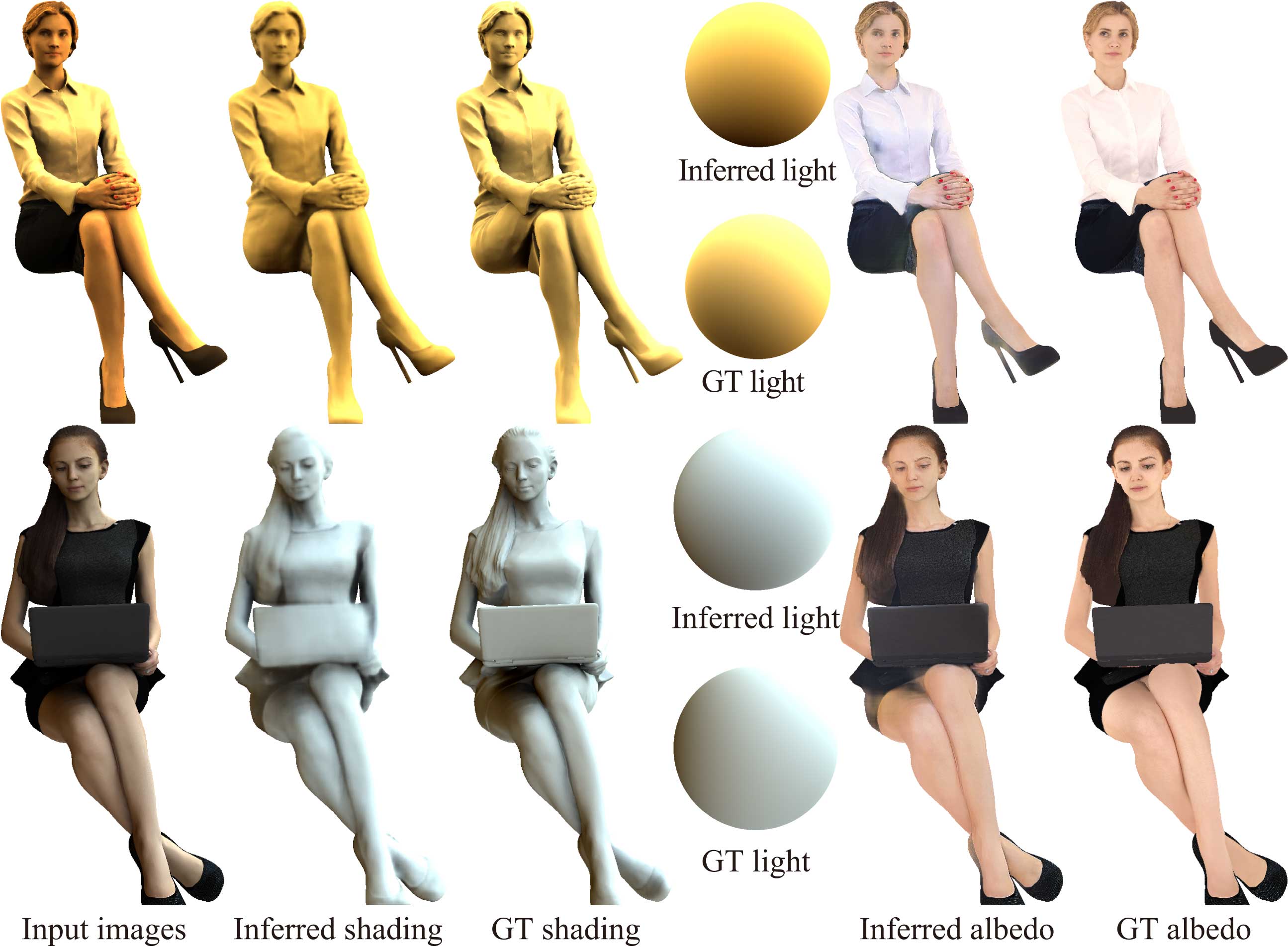}
  \caption{\rev{Inferred results with sitting poses. \revb{Our network can handle sitting poses unexpectedly well even though it is not trained with them.}}}
  \label{fig:SittingPoses}
 \end{figure}

\begin{figure}[t]
  \centering
  \includegraphics[width=\linewidth]{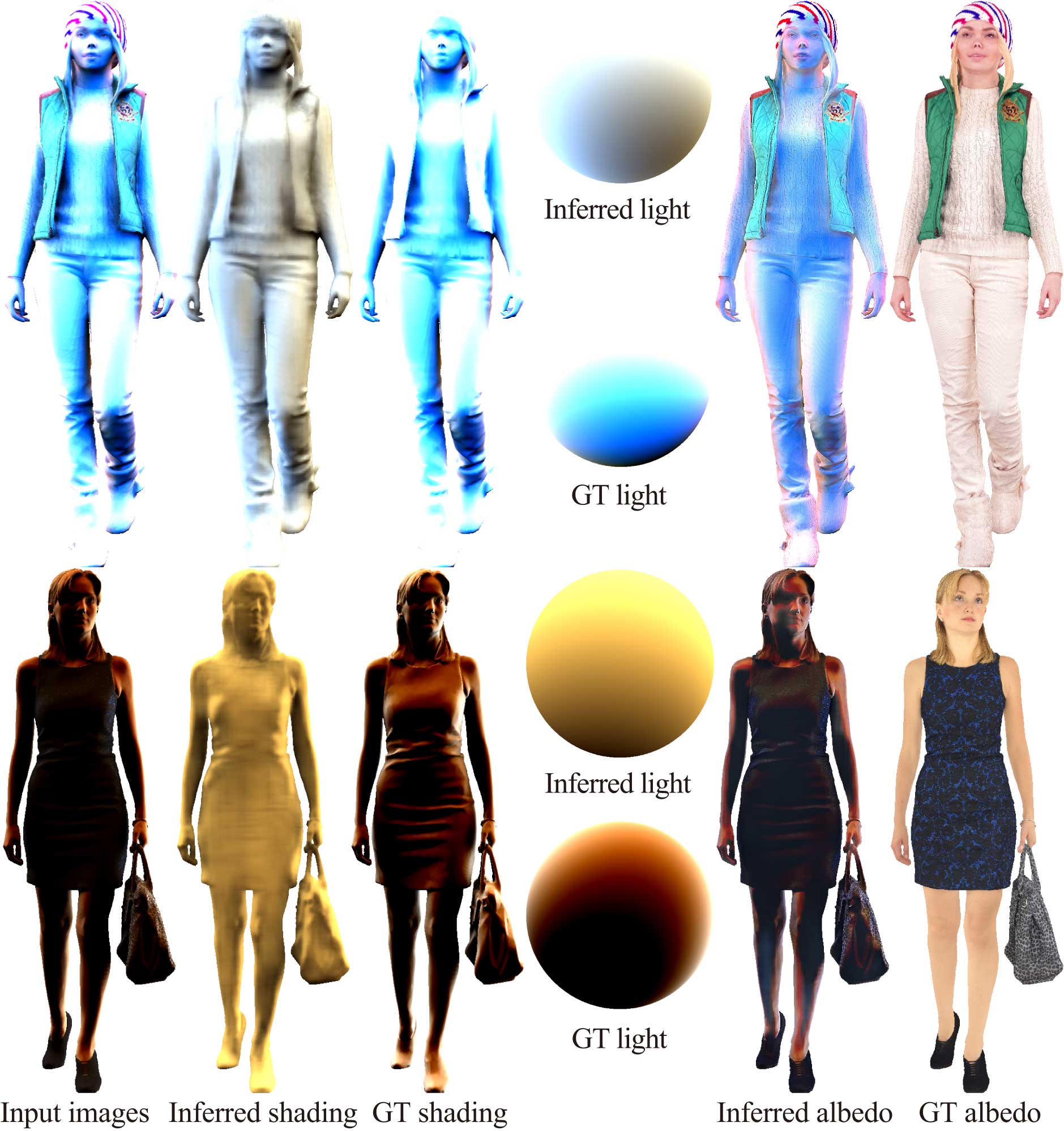}
  \caption{Failure examples with \revb{unusual} lights. \revb{Our network suffers from inferring lights quite different from the training data.}}
  \label{fig:LightFailure}
 \end{figure}

\paragraph*{Silhouettes as priors.}

Some existing methods~\cite{DBLP:journals/pami/BarronM15,Lun:2017:SketchModeling} on shape-from-shading suggested that object silhouettes serve as shape priors.
In the recent work of the CNN-based shape inference from 2D silhouettes~\cite{Lun:2017:SketchModeling}, the size of the training dataset is around ten thousand.
Compared to this size, it was a surprise that we can infer plausible albedo and shading from only a few hundreds of training data.
To confirm \revd{how} much the silhouettes help our inference, we inferred light transport maps only from the binary masks.
For this inference, we used the CNN model for light transport maps and used \revd{only those} loss functions related to light transport maps.
Figure~\ref{fig:ShapeFromMask} shows the resultant shading maps and corresponding ground-truth.
Surprisingly, we can observe the rough concave shapes under the chin \rev{and} the flat shapes of the instep.
This \revb{result} implies that our CNN models also learned a strong shape prior from silhouettes thanks to the regularity of our \revb{small} training dataset.

\paragraph*{\rev{Self-supervised learning.}}

Recent methods for intrinsic decomposition or image disentanglement, e.g., \cite{DBLP:conf/CVPR2018/abs-1712-01261}, often employ {\em self-supervised learning} to fine-tune networks that are trained with synthetic data;
\revb{only the single loss for the differences between input images and products of inferred outputs is considered, and the network is trained using unlabeled real photographs.
We fine-tuned the model of ``Ours (full)'' with and without fixing the network parameters of the decoders.
However, in both cases, the inferred outputs collapsed (see Figure~\ref{fig:SelfSupervisedLearning}); the light transport maps lost details, the corresponding shading maps bleached, and the albedo maps got close to the input images.}  
\revb{This is probably} because our light transport maps have much larger degrees of freedom (i.e., nine \revb{dimensions} per pixel) than normal maps inferred in \cite{DBLP:conf/CVPR2018/abs-1712-01261}, and thus are more difficult to fine-tune under the unconstrained setting in self-supervised learning.
We thus did not adopt self-supervised learning in other results.
The details of the experimental settings \revd{are} available in the supplemental material.

\paragraph*{\rev{Sitting poses.}}

To evaluate the ability of our network for handling various poses, we fed synthetic human images in sitting poses, which were not included in training or test data. 
Figure~\ref{fig:SittingPoses} shows the results.
The inferred outputs are unexpectedly well compared to the ground-truth, \revb{which} is probably because our training dataset is sufficiently rich for the network to learn shapes of body parts such as arms and legs. 

\paragraph*{\revb{Unusual} lights.}

We also evaluated the \revb{ability} for handling various lights, as shown in Figure~\ref{fig:LightFailure}.
Unfortunately, our network could not plausibly infer lights that were quite different from those in our training data.
Our network seems to reconstruct nearest-neighbor lights that can be found in the training dataset, and the light transport maps are inferred accordingly.
The large differences in appearance are then encoded in the inferred albedo maps so that the products of the three components become similar to the input images.
A straightforward solution is to enrich the training light dataset using, e.g., the environment maps used in \cite{DBLP:journals/tog/EndoKM17}, so that good nearest neighbors can be found for various inputs.


\subsection{Limitations}
\label{sec:Limitations}

Here we explain the limitations of our method.
Although our method is based on a better formulation of SH-based lighting, i.e., with consideration of light occlusion, it is still a crude approximation of lighting calculation.
First of all, we only handle diffuse albedo.
This \revb{limitation} mainly stems from our dataset; most of \revb{the} commercial data do not have specular components, even though SH representation can naturally handle specular components as demonstrated in the original PRT paper.
Adding artificial specular components to our training dataset, as done in \cite{DBLP:journals/cgf/InnamoratiRWM17}, seems inappropriate in our case because human skin and clothes should have different reflectance. 
Material assignment with semantic segmentation for hundreds of meshes is ideal \revd{but} can be a challenging project by itself.
As our work is the first attempt \revb{regarding} both full-body relighting and SH-based light occlusion learning/inference, we believe this limitation is acceptable to encourage follow-up studies.

Also, while we used \revb{second}-order SH for representing light occlusion, Sloan et al.~\shortcite{DBLP:journals/tog/SloanKS02} suggested to use higher-order bases because occlusion causes high-frequency signals.
As is often the case with learning-based methods, our method might fail with conditions quite dissimilar to the training dataset, e.g., harsh illuminations\rev{, as demonstrated in Figure~\ref{fig:LightFailure}}.


\section{Conclusions and Future Work}
\label{sec:Conclusions}

In this paper, we have paved the way to occlusion-aware relighting from single-view human images and accompanying inference using CNNs.
Inspired by the seminal work of the precomputed radiance transfer~\cite{DBLP:journals/tog/SloanKS02}, we employed SH-based lighting, i.e., dot-product calculation of \revb{second}-order spherical harmonics (SH) coefficient vectors of illumination and occlusion (i.e., light transfer vectors), and trained our models using our synthetic ground-truth dataset.
Plausible inference of albedo and light transport maps were possible probably because of our \revb{small} yet geometrically-aligned human image dataset.
By considering light occlusion, inferred albedo and shading maps (i.e., the product of a light transport map and illumination) as well as relighting results are more plausible than those \rev{obtained by} using \rev{previous} techniques without considering light occlusion.

One obvious direction of future work is to extend our first attempt to more physically-accurate inverse rendering, based on the formulations extensively studied in the literature of precomputed radiance transfer.
For example, other basis functions such as wavelets or spherical Gaussians might be beneficial to handle high-frequency shadows or illumination.
A quite important future work would be to build a publicly-available, high-quality 3D human models, which is crucial to develop this human-oriented research.

\begin{acks}
The authors would like to thank ZOZO Technologies, Inc. for generous financial support throughout this project, without which this work was not possible.  
The authors would also like to thank the anonymous referees for their constructive comments, and Ms. Sina Kitz for proof-reading the final version of this paper.
For our accompanying video, input images courtesy of Kat Garcia, Kinga Cichewicz, George Gvasalia, and Jacob Postuma.
\end{acks}


\bibliographystyle{ACM-Reference-Format}
\bibliography{relighting_selfies}
\end{document}